\documentclass[submission,copyright,creativecommons]{eptcs}
\providecommand{\event}{GASCOM 2026}

\usepackage{iftex}
\usepackage[utf8]{inputenc}
\usepackage[english]{babel}
\usepackage{xcolor}
\usepackage{amsmath}
\usepackage{amsthm}
\usepackage{enumerate}
\usepackage[warn]{textcomp}
\usepackage{graphicx}
\usepackage{caption}
\usepackage{tikz}
\usepackage{hyperref}
\usepackage[normalem]{ulem}

\definecolor{myred}{rgb}{1,0,0}
\definecolor{myorange}{rgb}{1,0.5,0}

\newtheorem{theorem}{Theorem}

\newtheorem{definition}[theorem]{Definition}

\newtheorem{remark}[theorem]{Remark}

\ifpdf
  \usepackage[T1]{fontenc}
\else
  \usepackage{breakurl}
\fi
\title{
A Class of \\Multiparameter Signless Stirling Numbers of the First Kind and their 
$q$-Analogues
}
\author{Violetta E. Piperigou
\institute{Department of Mathematics\\
University of Patras, Greece
}
\email{vpiperig@math.upatras.gr}
\and
Malvina G. Vamvakari
\institute{Department of Informatics $\&$ Telematics\\
Harokopio University of Athens, Greece}
\email{mvamv@hua.gr
}}
\def\titlerunning{
Multiparameter 
Signless Stirling Numbers of the First Kind and 
$q$-Analogues
 }
\def\authorrunning{V. E. Piperigou \& M.G. Vamvakari
}

\begin{document}
\maketitle

\begin{abstract}
In this work, we provide a probabilistic derivation of a class of multiparameter signless Stirling numbers of the first kind and their $q$-analogues, and study the associated multivariate discrete distributions.

\end{abstract}
\section{Brief Introduction}
The present work is motivated by earlier contributions of Cacoullos and Papageorgiou \cite{CacoulPap} and Koutras \cite{Koutras1990}, as well as more recent developments by Charalambides \cite{Charal1,Charal2,Charal3,Charal4}, Kyriakoussis and Vamvakari \cite{kyrvam} and Vamvakari \cite{malvina}. Inspired by these works, we introduce a class of multiparameter Stirling numbers and their $q$-analogues through probabilistic models.
\newline
In the first part, we extend to the multivariate setting the univariate Stirling distributions from which the corresponding signless Stirling numbers of the first kind arise, as presented in Charalambides \cite{Charal1}, together with the associated class of multiparameter signless Stirling numbers of the first kind. In the second part, we proceed analogously by generalizing the univariate discrete $q$-Stirling distributions presented in Charalambides \cite{Charal2} to their multivariate counterparts, and derive the corresponding class of multiparameter signless $q$-Stirling numbers of the first kind.
\newline
Our approach provides a  probabilistic framework for the study of these classes of multiparameter signless Stirling numbers of the first kind and their 
$q$-analogues, and leads naturally to the establishment of the associated multivariate discrete distributions.
%\newline
\section {Main Results}
\subsection{A Probability Model for a Class of
Multiparameter Signless Stirling Numbers of the First Kind}\label{subsec}
 Consider a chamber of $k$ consecutive lined cells initially containing $m_j$ balls of color $c_j$, where $c_j$ denotes the $j$-th shade of grey, for $j=1,2,\ldots,k, $ and a group of $n$ batches of $s$ red balls.
 %\newline
 Suppose that $n$ balls are successively drawn one after the other  from the first lined  cell  where after each trial the drawn ball is placed back in the first lined cell  along with $s$ red balls.
\newline
Let $A_{1,i}$ be the event of drawing a ball of color $c_1$
%$\textcolor[gray]{0.25}{c_1}$
 at the $i$-th trial, $i=1,2,\ldots,n$. Then, setting $\theta_1=m_1/s$, we get
\begin{eqnarray*}
p_{1,i}=P\left(A_{1,i}\right)=\frac{\theta_1}{\theta_{1}+i-1}, \,\, p'_{1,i}=P\left(A'_{1,i}\right)=\frac{i-1}{\theta_{1}+i-1}, i=1,2,\ldots,n, 
\end{eqnarray*}
where $A'_{1,i}$ denotes the complementary  event of  drawing a ball of color red at the $i$-th trial, $i=1,2,\ldots,n$. 
After $n$ drawings (trials) we remove from the first cell $n-x_1$ batches of $s$ red balls, where $x_1$ is the number of balls of color $c_1$
 %$\textcolor[gray]{0.25}{c_1}$ 
 drawn and placed back to the first cell, $x_1=1,2,\ldots,n$. 
%%{Drawings: 2$^{\mbox{nd}}$ cell}
%Group of $n-x_1$, $x_1=1,2,\ldots,n,$  batches of $s$ red balls 
Now, suppose that 
%Successively 
$n-x_1$ balls are successively drawn one after the other from the second lined cell 
% drawings, 
 where after each trial the drawn ball is placed back along with $s$ red balls.
 \newline
Let $A_{2,i}$ be the event of drawing a ball of color $c_2$
% $\textcolor[gray]{0.45}{c_2}$
  at the $i$-th trial, $i=1,2,\ldots,n-x_1$. Then, setting $\theta_2=m_2/s$, we get
\begin{eqnarray*}
&&p_{2,i}=P\left(A_{2,i}\right)=\frac{\theta_2}{\theta_{2}+i-1},p'_{2,i}=P\left(A'_{2,i}\right)=\frac{i-1}{\theta_{2}+i-1},\\ \\
 &&\,\,\,\,\,\,i=1,2,\ldots,n-x_1,\,\,\,x_1=1,2,\ldots,n.
\end{eqnarray*}
After $n-x_1$ drawings (trials) we remove from the second cell $n-x_1-x_2$ batches of $s$ red balls, where $x_2$ is the number of balls of color $c_2$
% $\textcolor[gray]{0.45}{c_2}$ 
 drawn and placed back to the second cell, $x_2=1,2,\ldots,n-x_1$. 
 \newline
 As regards the drawings from the $ k$-th cell, there is a group of $n-\sum_{j=0}^{k-1}x_j$ batches of $s$ red balls.
 % ($x_0$=0). 
% {Drawings: k$^{\mbox{th}}$ cell}
% Group of $n-\sum_{j=0}^{k-1}x_j$ batches of $s$ red balls ($x_0$=0)
Suppose that  we successively  draw $n-\sum_{j=0}^{k-1}x_j$ balls from the $k$-th cell, where after each trial the drawn ball is placed back along with $s$ red balls.
Let $A_{k,i}$ be the event of drawing a ball of color $c_k$
% $\textcolor[gray]{0.8}{c_k}$ 
 at the $i$-th trial, $i=1,2,\ldots,n-\sum_{j=0}^{k-1}x_j$. Then, setting $\theta_k=m_k/s$, we get
\begin{eqnarray*}
p_{k,i}=P\left(A_{k,i}\right)=\frac{\theta_k}{\theta_{k}+i-1},p'_{k,i}=P\left(A'_{k,i}\right)=\frac{i-1}{\theta_{k}+i-1},%\\
 \end{eqnarray*}
$i=1,2,\ldots,n-\sum_{j=0}^{k-1}x_j,$  $x_k=1,2,\ldots,n-\sum_{j=0}^{k-1}x_j$, $\,k \geq 1$,  where $x_k$ are the number of balls of color $c_k$
% $\textcolor[gray]{0.8}{c_k}$
  drawn  and placed back to the $k$-th cell in $n-\sum_{j=0}^{k-1}x_j$ trials, with $x_0=0$. % from the $k^{th}$ cell. %,\,\,\,x_1=1,2,\ldots,n.
So,
\begin{eqnarray*}
&&P\left( \prod_{j=1}^k A_{j,i_{j,1}}A_{j,i_{j,2}}\ldots A_{j,i_{j,x_j}}A'_{j,i_{j,x_j+1}}\ldots A'_{j,i_{j,n-\sum_{\nu=0}^{j-1}x_\nu}} \right)\\
&&=\prod_{j=1}^k P\left(A_{j,i_{j,1}}\right)
P\left(A_{j,i_{j,2}}\right)\ldots P\left(A_{j,i_{j,x_j}}\right) \\
&&\hspace*{1cm}\cdot P\left(A'_{j,i_{j,x_j+1}}\right) \ldots P\left(A'_{j,i_{j,n-\sum_{\nu=0}^{j-1}x_\nu}}\right) 
\end{eqnarray*}
Therefore, the probability $p\left(x_1,x_2,\ldots,x_k;n\right)$ of drawing $x_j$ balls of color $c_j$ from the $j$-th cell, $j=1,2,\ldots,k$, $k \geq 1$, under the described probability model, is given by the equation
\begin{eqnarray*}
&&p\left(x_1,x_2,\ldots,x_k;n\right)=\frac{\prod_{j=1}^k {\theta_j}^{x_j}}{\prod_{j=1}^k\left(\theta_j+n-\sum_{\nu=0}^{j-1}x_\nu-1\right)_{n-\sum_{\nu=0}^{j-1}x_\nu}}\\
&&\,\,\,\cdot \sum \left(i_{1,x_1+1}-1\right)\cdots  \left(i_{1,n}-1\right)\left(i_{2,x_2+1}-1\right)\cdots  \left(i_{2,n-x_1}-1\right)\\
&&\,\,\,\,\,\,\,\,\,\,\,\,\,\,\,\,\,\,\cdots\left(i_{k,x_k+1}-1\right)\cdots  \left(i_{1,n-\sum_{j=0}^{k-1}x_j}-1\right), \sum_{j=1}^kx_j \leq n, k \geq 1, \\
&&\,\,\,\,\,\,\,\,\,\,\,\,\,\,\,\,\,\, \,\,\,\,x_j \geq1, j=1,2,\dots,k, x_0=0, 
\end{eqnarray*}
where the multiple sum is extended over all $\left(n-\sum_{\nu=1}^j x_\nu\right)$-combinations $\left \{i_{j,x_j+1},i_{j,x_j+2},\ldots,i_{j,n-\sum_{\nu=0}^{j-1}x_\nu}\right \}$ of the $n-\sum_{\nu=0}^{j-1}x_\nu-1$ positive integers $\left \{2,3,\ldots,n-\sum_{\nu=0}^{j-1}x_\nu \right \},$ $j=1,2,\ldots,k$, $k \geq 1$.
Thus,
\begin{eqnarray*}
&&p\left(x_1,x_2,\ldots,x_k;n\right)=\frac{\prod_{j=1}^k {\theta_j}^{x_j}}{\prod_{j=1}^k\left(\theta_j+n-\sum_{\nu=0}^{j-1}x_\nu-1\right)_{n-\sum_{\nu=0}^{j-1}x_\nu}}\\
&&\,\,\,\cdot \sum m_{1,1}\cdots  m_{1,n-x_1}m_{2,1}\cdots  m_{2,n-x_1-x_2}\cdots m_{k,1}\cdots  m_{k,n-\sum_{\nu=1}^kx_\nu},  \\
&&\,\,\,\,\,\,\,\,\,\,\,\,\,\,\,\,\,\,
%\cdots\left(i_{k,x_k+1}-1\right)\cdots  \left(i_{1,n-\sum_{j=0}^{k-1}x_j}-1\right),
 \sum_{j=1}^kx_j \leq n, k \geq 1, x_j \geq1, j=1,2,\dots,k, x_0=0, % \nonumber\\
%&&\,\,\,\,\,\,\,\,\,\,\,\,\,\,\,\,\,\, \,\,\,\,
\end{eqnarray*}
where the multiple sum is extended over all $\left(n-\sum_{\nu=1}^j x_\nu\right)$-combinations $\left \{m_{j,1},m_{j,2},\ldots,m_{j,n-\sum_{\nu=1}^{j}x_\nu}\right \}$ of the $n-\sum_{\nu=0}^{j-1}x_\nu-1$ positive integers $\left \{1,2,3,\ldots,n-\sum_{\nu=0}^{j-1}x_\nu -1\right \},$ $j=1,2,\ldots,k$, $k \geq 1$.
\begin{definition}
\label{def1}
Let $|s\left(n,x_1,x_2,\ldots,x_k\right)|,$  $\sum_{j=1}^kx_j \leq n,$ $x_j\geq 1$, $j=1,2\ldots,k,$ $k \geq 1$, be the numbers given by the multiple sum 
\begin{eqnarray*}
|s\left(n,x_1,x_2,\ldots,x_k\right)|
=\sum \prod_{j=1}^km_{j,1}\cdots  m_{j,n-\sum_{\nu=1}^jx_\nu}
%m_{2,1}\cdots  m_{2,n-x_1-x_2}\cdots m_{k,1}\cdots  m_{k,n-\sum_{\nu=1}^kx_\nu}
,
\end{eqnarray*}
where the multiple sum is extended over all $\left(n-\sum_{\nu=1}^j x_\nu\right)$-combinations $\left \{m_{j,1},m_{j,2},\ldots,m_{j,n-\sum_{\nu=1}^{j}x_\nu}\right \}$ of the %\textcolor{red}{
$n-\sum_{\nu=0}^{j-1}x_\nu-1$ positive integers $\left \{1,2,3,\ldots,n-\sum_{\nu=0}^{j-1}x_\nu -1\right \},$
%}
 $j=1,2,\ldots,k$, $k \geq 1$. The numbers  $|s\left(n,x_1,x_2,\ldots,x_k\right)|$ are called %\emph{signless \textcolor{red}{
multiparameter signless
 %}
  Stirling numbers of the first kind.
\end{definition}
%\end{frame}
%\begin{frame}
\begin{remark}
%\
For $k=1,$ the numbers $|s(n,x_1)|,$ $x_1=1,2,\ldots,n,$ according to the previous definition, are %the  signless Stirling numbers of the first kind, 
 given by
\begin{equation*}
|s(n,x_1)|=\sum m_{1,1}\cdots  m_{1,n-x_1}
\end{equation*}
where the summation is extended over all $\left(n-x_1\right)$-combinations $\left \{m_{1,1},m_{1,2},\ldots,m_{1,n-x_1}\right \}$ of the $n-1$ positive integers $\left \{1,2,3,\ldots,n-1\right \}$. Therefore, $|s(n,x_1)|$ are the signless Stirling numbers of the first kind.
\end{remark}
\subsection{A Probability Model for Multiparameter Signless Noncentral Stirling Numbers of the First Kind}

Next, we 
%expand the previous probability model by
 consider a  chamber of $k$ consecutive lined cells initially containing $m_j$ balls of color $c_{1:j}$ and $b_j$ balls of color $c_{2:j}$, $j=1,2,\ldots,k$ %$\bluebullet$
and a
group of $n$ batches of $s$ red balls. Suppose that we apply the previous probability model of successive drawings described in Section \ref{subsec}.
%\newline
 Let
$A_{j,i}$ be the event of drawing a ball of color  $c_{1:j},$ $j=1,2,\ldots,k$, $k \geq 1,$
 %$\textcolor[gray]{0.8}{c_k}$ 
at the $i$-th trial, $i=1,2,\ldots,n-\sum_{\nu=0}^{j-1}x_j$. 
Then, 
 setting $\theta_j=m_j/s$, $r_j=b_j/s$, we get
\begin{eqnarray*}
p_{j,i}=P\left(A_{j,i}\right)=\frac{\theta_j}{\theta_{j}+r_j+i}, \,\, p'_{j,i}=P\left(A'_{j,i}\right)=\frac{r_j+i}{\theta_{j}+r_j+i}, 
%i=1,2,\ldots,n 
\end{eqnarray*}
$i=1,2,\ldots,n-\sum_{\nu=0}^{j-1}x_\nu,$  $x_j=0,1,2,\ldots,n-\sum_{\nu=0}^{j-1}x_\nu$, where $x_j$ are the number of balls of color $c_{1:j}$ drawn  and placed back to the $j$-th cell in $n-\sum_{\nu=0}^{j-1}x_\nu$, trials $j=1,2,\ldots,k$.
\newline
Working analogously as in 
%the  probability model of 
Section \ref{subsec}, 
%we expand definition \ref{def1} and 
we  derive the following expansion of Definition \ref{def1}.
\begin{definition}
Let $|s\left(n,x_1,x_2,\ldots,x_k;r_1,r_2,\ldots,r_k\right)|,$  $\sum_{j=1}^kx_j \leq n,$ $x_j\geq 1$, $j=1,2\ldots,k,$ $k \geq 1$, be the numbers given by the multiple sum 
\begin{align*}
&|s\left(n,x_1,x_2,\ldots,x_k;r_1,r_2,\ldots,r_k\right)|\\
&=\sum \prod_{j=1}^k\left(r_1+m_{j,1}\right)\cdots  \left(r_j+m_{j,n-\sum_{\nu=1}^jx_\nu}\right)
%m_{2,1}\cdots  m_{2,n-x_1-x_2}\cdots m_{k,1}\cdots  m_{k,n-\sum_{\nu=1}^kx_\nu}
,
\end{align*}
\newline
where the multiple sum is extended over all $\left(n-\sum_{\nu=1}^j x_\nu\right)$-combinations $\left \{m_{j,1},m_{j,2},\ldots,m_{j,n-\sum_{\nu=1}^{j}x_\nu}\right \}$ of the $n-\sum_{\nu=0}^{j-1}x_\nu-1$ positive integers $\left \{1,2,3,\ldots,n-\sum_{\nu=0}^{j-1}x_\nu -1\right \},$ $j=1,2,\ldots,k$, $k \geq 1$. 
The numbers  $|s\left(n,x_1,x_2,\ldots,x_k;r_1,r_2,\ldots,r_k\right)|$ are called multiparameter signless  noncentral Stirling numbers of the first kind.
\end{definition}
\begin{remark}
%\
For $k=1,$ the numbers $|s(n,x_1;r_1)|,$ $x_1=1,2,\ldots,n,$ according to the previous definition, are %the  signless Stirling numbers of the first kind, 
 given by
\begin{equation*}
|s(n,x_1;r_1)|=\sum \left(r_1+m_{1,1}\right)\cdots \left(r_1+ m_{1,n-x_1}\right)
\end{equation*}
where the summation is extended over all $\left(n-x_1\right)$-combinations $\left \{m_{1,1},m_{1,2},\ldots,m_{1,n-x_1}\right \}$ of the $n-1$ positive integers $\left \{1,2,3,\ldots,n-1\right \}$. Therefore, $|s(n,x_1;r_1)|$ are the signless noncentral Stirling numbers of the first kind (see Charalambides \cite{Charal1}).

\end{remark}

%%%%%%%%%%%%%%%%%%%%%%%%%%%
 \subsection{A Probability Model for a Class of Multiparameter Signless 
 %Noncentral 
 $q$-Stirling Numbers of the First Kind}

The multivariate discrete  $q$-distributions, $0<q<1$, are based on  stochastic models of  sequences of $n$ independent Bernoulli trials with chain-composite failures (or successes),
 % is considered, 
 where the odds of success of a certain kind at a trial is assumed to vary geometrically, with rate $q$, with the number of previous trials or with the number of previous successes or both with  the number of  previous trials and
 successes (see Charalambides \cite{Charal5}).
 \newline
Next, we consider a  sequence of %$n$ 
  independent Bernoulli trials with chain-composite failures,
  where the
%  \item
   probability of success  of the $j$-th kind at the $i$-th trial is given by 
  \begin{eqnarray}
  \label{sucess}
  p_{j,i}=\frac{1}{[r_j+i]_q},\,\,
 % \,\, \mbox{and} \,\, p'_{j,i}=q\frac{[r_j+i-1]_q}{[r_j+i]_q},
    0\leq r_j<\infty,\,\, j=1,2,\ldots,k, i=1,2,\ldots
    %,n
    %,\,0<q<1 %\nonumber
  \end{eqnarray}
% where $0<q<1$. 
%\item
  Let $A_{j,i}$ be the event of success of the $j$-th kind at the $i$-th trial with probability of success given by Eq.~\eqref{sucess} for $j=1,\ldots,k$, $k \geq 1$, $i=1,\ldots,%, n
   $ % Then, using the i
     and consider a permutation $(i_{j,1},\ldots,i_{j,x_j},i_{j,x_{j+1}},\ldots, i_{j,n-\sum_{\nu=0}^{j-1}x_\nu})$ of 
  $\{1,2,\ldots, n-\sum_{\nu=0}^{j-1}x_{\nu} \}$. Then, we get
%   \item 
%From the    independence of Bernoulli trials we have that
  % \item
%  \vspace{-0.4cm}
  \begin{eqnarray*}
&&P\left( \prod_{j=1}^k A_{j,i_{j,1}}%A_{j,i_{j,2}}
\ldots A_{j,i_{j,x_j}}A'_{j,i_{j,x_j+1}}\cdots A'_{j,i_{j,n-\sum_{\nu=0}^{j-1}x_\nu}} \right)\\
%\nonumber\\
&&=\prod_{j=1}^k P\left(A_{j,i_{j,1}}\right) 
%%P\left(A_{j,i_{j,2}}\right)
\ldots P\left(A_{j,i_{j,x_j}}\right)% \nonumber\\
%&&\hspace*{1cm}\cdot 
P\left(A'_{j,i_{j,x_j+1}}\right) \cdots P\left(A'_{j,i_{j,n-\sum_{\nu=0}^{j-1}x_\nu}}\right) 
\end{eqnarray*} and the following theorem holds.
%\pagebreak
 \begin{theorem}
 Let $X_j$ be the number of successes of the $j$-th kind in a 
   sequence of $n$ independent Bernoulli trials with chain-composite failures, where the probability of success  of the $j$-th kind at the $i$-th trial is given by Eq.~\eqref{sucess}, for $j=1,2,\ldots,k$. Then the probability function of the r.v.  $\left (X_1,X_2,\ldots,X_k \right) $ is given by
   \begin{eqnarray*}
   P\left (X_1=x_1,X_2=x_2,\ldots,X_k=x_k\right)&=&
   %\prod_{j=1}^k\dfrac{q^{n-\sum_{\nu=1}^jx_\nu}}{\,[r_j+n-\sum_{\nu=0}^{j-1}x_\nu]_{n-\sum_{\nu=0}^{j-1}x_\nu}} 
   \dfrac{q^{nk-\sum_{j=1}^k(k-j+1)x_j}}{\prod_{j=1}^k[r_j+n-\sum_{\nu=0}^{j-1}x_\nu]_{{n-\sum_{\nu=0}^{j-1}x_\nu},q}} \\
   && \cdot \sum \prod_{j=1}^k [r_j+m_{j,1}]_q[r_j+m_{j,2}]_q\cdots[r_j+m_{j,n-\sum_{\nu=1}^jx_\nu}]_q, x_j=0,1,\ldots,n,%\nonumber\\
   \end{eqnarray*}
  % \vspace{-0.5em}
%for $x_j=0,1,\ldots,n,$ 
with  $\sum_{j=1}^kx_j \leq n$, $x_0=0$, 
where the multiple sum is extended over all $\left(n-\sum_{\nu=1}^j x_\nu\right)$-combinations $\left \{m_{j,1},m_{j,2},\ldots,m_{j,n-\sum_{\nu=1}^{j}x_\nu}\right \}$ of the $n-\sum_{\nu=0}^{j-1}x_\nu$ nonnegative integers $\left \{0,1,\ldots,n-\sum_{\nu=0}^{j-1}x_\nu -1\right \},$ $j=1,2,\ldots,k$, $k \geq 1$.

  \end{theorem}

\begin{definition}

Let $|s_q\left(n,x_1,x_2,\ldots,x_k;r_1,r_2,\ldots,r_k\right)|,$ $\sum_{j=1}^kx_j \leq n,$ $x_j\geq 1$, $0\leq r_j<\infty$, $j=1,2\ldots,k,$ $k \geq 1$, be the numbers given by the multiple sum 
\begin{eqnarray*}
%&&
&&|s_q\left(n,x_1,x_2,\ldots,x_k;r_1,r_2,\ldots,r_k\right)|=q^{nk-\sum_{j=1}^k(k-j+1)x_j}\\
%&&\,\,\,\,\,\,\,\,\,
&&\cdot \sum \prod_{j=1}^k [r_j+m_{j,1}]_q[r_j+m_{j,2}]_q\cdots[r_j+m_{j,n-\sum_{\nu=1}^jx_\nu}]_q,%\, x_j=0,1,\ldots,n
\end{eqnarray*}
where the multiple sum is extended over all $\left(n-\sum_{\nu=1}^j x_\nu\right)$-combinations $\left \{m_{j,1},m_{j,2},\ldots,m_{j,n-\sum_{\nu=1}^{j}x_\nu}\right \}$
of the $n-\sum_{\nu=0}^{j-1}x_\nu$ nonnegative integers 
$\left \{0,1,\ldots,n-\sum_{\nu=0}^{j-1}x_\nu -1\right \}$
, $j=1,2,\ldots,k$, $k \geq 1$. 
The numbers  $|s_q\left(n,x_1,x_2,\ldots,x_k;r_1,r_2,\ldots,r_k\right)|$ are called multiparameter signless noncentral $q$-Stirling numbers of the first kind.
\end{definition}
%\end{frame}
%\begin{frame}
\begin{remark}

For $k=1,$ the numbers $|s_q(n,x_1;r_1)|,$ $x_1=1,2,\ldots,n,$ $n=1,2,\ldots,$ $0 \leq r_1 <\infty,$ according to the previous definition, are %the  signless Stirling numbers of the first kind, 
 given by
\begin{equation}
|s_q(n,x_1;r_1)|=%\sum m_{1,1}\cdots  m_{1,n-x_1}\nonumber
q^{n-x_1}%\nonumber\\
%&&\,\,\,\,\,\,\,\,\,
%&&\cdot 
\sum  [r_1+m_{1,1}]_q[r_1+m_{1,2}]_q\cdots[r_1+m_{1,n-x_1}]_q,
\end{equation}
where the summation is extended over all $\left(n-x_1\right)$-combinations $\left \{m_{1,1},m_{1,2},\ldots,m_{1,n-x_1}\right \}$ of the $n$ nonnegative integers $\left \{0,1,\ldots,n-1\right \}$. Therefore, $|s_q(n,x_1;r_1)|$ are the  signless noncentral $q$-Stirling numbers of the first kind (see Charalambides \cite{Charal2}).
\end{remark}
 
 \begin{remark}
 The multiparameter noncentral $q$-Stirling numbers of the first kind %$s_q\left(n,x_1,x_2,\ldots,x_k;r_1,r_2,\ldots,r_k\right),$ 
 are defined by the multiple sum as follows
 \begin{eqnarray*}
%&&
&&s_q\left(n,x_1,x_2,\ldots,x_k;r_1,r_2,\ldots,r_k\right)%q^{nk-\sum_{j=1}^k(k-j+1)x_j}
\nonumber\\
%&&\,\,\,\,\,\,\,\,\,
%&&\cdot
&&= \sum \prod_{j=1}^k [-1]_q^{n-\sum_{\nu=1}^{j}x_\nu} [r_j+m_{j,1}]_q[r_j+m_{j,2}]_q\cdots[r_j+m_{j,n-\sum_{\nu=1}^jx_\nu}]_q,%\, x_j=0,1,\ldots,n
%\sum \prod_{j=1}^k\prod_{\nu_j=x_j+1}^{n-\sum_{\nu=0}^{j-1}x_\nu}
  % [r_j+i_{j,\nu_j}-1]_q
%\sum m_{1,1}\cdots  m_{1,n-x_1}m_{2,1}\cdots  m_{2,n-x_1-x_2}\cdots m_{k,1}\cdots  m_{k,n-\sum_{\nu=1}^kx_\nu}
\nonumber
\end{eqnarray*}
where the summation is extended over all $\left(n-\sum_{\nu=1}^j x_\nu\right)$-combinations $\left \{m_{j,1},m_{j,2},\ldots,m_{j,n-\sum_{\nu=1}^{j}x_\nu}\right \}$ of the $n-\sum_{\nu=0}^{j-1}x_\nu$ nonnegative integers $\left \{0,1,\ldots,n-\sum_{\nu=0}^{j-1}x_\nu -1\right \},$ $j=1,2,\ldots,k$, $k \geq 1$.
\end{remark}
 \noindent
 Note that if we change suitably the probability of success the following theorem also holds.
 \begin{theorem}
 Let $X_j$ be the number of successes of the $j$-th kind in a 
   a sequence of $n$ independent Bernoulli trials with chain-composite failures, where the probability of success  of the $j$-th kind at the $i$-th trial, is given by 
 %  (\ref{sucess}), for $j=1,2,\ldots,k$. 
  \begin{eqnarray}
  \label{sucess2}
  p_{j,i}=q^{r_j+i-1},\,\,
 % \,\, \mbox{and} \,\, p'_{j,i}=q\frac{[r_j+i-1]_q}{[r_j+i]_q},
    0\leq r_j<\infty,\,\, j=1,2,\ldots,k, i=1,2,\ldots,
  %  n,
    \,0<q<1. 
  \end{eqnarray} 
   Then the probability function of the r.v.  $\left (X_1,X_2,\ldots,X_k \right) $ is given by
   \begin{eqnarray*}
  P\left (X_1=x_1,X_2=x_2,\ldots,X_k=x_k\right)%\nonumber\\
  &=&q^{\sum_{j=1}^k {n-\sum_{\nu=0}^{j-1}x_\nu \choose 2}+\sum_{j=1}^k \left(n-\sum_{\nu=0}^{j-1} x_\nu \right)
   r_j}\\&&\,\,\,\,\cdot 
   (1-q)^{\sum_{j=1}^k \left(n-\sum_{\nu=1}^{j}x_\nu\right)}
%   \nonumber\\
%   &\cdot
   |s_{q^{-1}}\left(n,x_1,x_2,\ldots,x_k;r_1,r_2,\ldots,r_k\right)|
\end{eqnarray*}
  % \vspace{-0.5em}
%for $x_j=0,1,\ldots,n,$ 
with $x_j=0,1,\ldots,n,$ $\sum_{j=1}^kx_j \leq n$, $x_0=0$,  where
\begin{eqnarray*}
|s_{q^{-1}}\left(n,x_1,x_2,\ldots,x_k;r_1,r_2,\ldots,r_k\right)|&=&(-1)^{\sum_{j=1}^k \left(n-\sum_{\nu=1}^{j}x_\nu\right)} \\&&\,\,\,\,\,\,\,\cdot
% (-1)^{\sum_{j=1}^k \left(n-\sum_{\nu=1}^{j}x_\nu\right)} 
 q^{-{\sum_{j=1}^k \left(n-\sum_{\nu=1}^{j}x_\nu\right)}}s_{q^{-1}}\left(n,x_1,x_2,\ldots,x_k;r_1,r_2,\ldots,r_k\right). 
\end{eqnarray*}

 \end{theorem}

\nocite{*}
\bibliographystyle{eptcs}
\bibliography{Vamvakari3}

\begin{thebibliography}{1}
\providecommand{\bibitemdeclare}[2]{}
\providecommand{\surnamestart}{}
\providecommand{\surnameend}{}
\providecommand{\urlprefix}{Available at }
\providecommand{\url}[1]{\texttt{#1}}
\providecommand{\href}[2]{\texttt{#2}}
\providecommand{\urlalt}[2]{\href{#1}{#2}}
\providecommand{\doi}[1]{doi:\urlalt{https://doi.org/#1}{#1}}
\providecommand{\eprint}[1]{arXiv:\urlalt{https://arxiv.org/abs/#1}{#1}}
\providecommand{\bibinfo}[2]{#2}

\bibitemdeclare{article}{CacoulPap}
\bibitem{CacoulPap}
\bibinfo{author}{Theophilos \surnamestart Cacoullos\surnameend} \&
  \bibinfo{author}{Haralambos \surnamestart Papageorgiou\surnameend}
  (\bibinfo{year}{1984}): \emph{\bibinfo{title}{Multiparameter Stirling and
  C-Numbers: Recurrences and Applications}}.
\newblock {\slshape \bibinfo{journal}{Fibonacci Quarterly}}
  \bibinfo{volume}{22}(\bibinfo{number}{2}), pp. \bibinfo{pages}{119--133},
  \doi{10.1080/00150517.1984.12429903}.

\bibitemdeclare{book}{Charal1}
\bibitem{Charal1}
\bibinfo{author}{Charalambos~A. \surnamestart Charalambides\surnameend}
  (\bibinfo{year}{2002}): \emph{\bibinfo{title}{Enumerative Combinatorics}}.
\newblock \bibinfo{publisher}{Chapman \& Hall/CRC}, \bibinfo{address}{Boca
  Raton, Florida}, \doi{10.1201/9781315273112}.

\bibitemdeclare{book}{Charal2}
\bibitem{Charal2}
\bibinfo{author}{Charalambos~A. \surnamestart Charalambides\surnameend}
  (\bibinfo{year}{2016}): \emph{\bibinfo{title}{Discrete $q$-Distributions}}.
\newblock \bibinfo{publisher}{John Wiley $\&$ Sons}, \bibinfo{address}{Hoboken,
  NJ}, \doi{10.1002/9781119119128}.

\bibitemdeclare{article}{Charal3}
\bibitem{Charal3}
\bibinfo{author}{Charalambos~A. \surnamestart Charalambides\surnameend}
  (\bibinfo{year}{2021}): \emph{\bibinfo{title}{$q$-Multinomial and Negative
  $q$-Multinomial Distributions}}.
\newblock {\slshape \bibinfo{journal}{Communications in Statistics - Theory and
  Methods}} \bibinfo{volume}{50}, pp. \bibinfo{pages}{5673--5898},
  \doi{10.1080/03610926.2020.1737711}.

\bibitemdeclare{article}{Charal4}
\bibitem{Charal4}
\bibinfo{author}{Charalambos~A. \surnamestart Charalambides\surnameend}
  (\bibinfo{year}{2022}): \emph{\bibinfo{title}{Multivariate $q$-Pólya and
  Inverse $q$-Pólya Distributions}}.
\newblock {\slshape \bibinfo{journal}{Communications in Statistics - Theory and
  Methods}} \bibinfo{volume}{51}, pp. \bibinfo{pages}{4854--4876},
  \doi{10.1080/03610926.2020.1825740}.

\bibitemdeclare{book}{Charal5}
\bibitem{Charal5}
\bibinfo{author}{Charalambos~A. \surnamestart Charalambides\surnameend}
  (\bibinfo{year}{2024}): \emph{\bibinfo{title}{Multivariate Discrete
  $q$-Distributions}}.
\newblock \bibinfo{publisher}{Springer Nature}, \bibinfo{address}{Switzerland
  AG}, \doi{10.1007/978-3-031-43713-7}.

\bibitemdeclare{article}{Koutras1990}
\bibitem{Koutras1990}
\bibinfo{author}{Markos~V. \surnamestart Koutras\surnameend}
  (\bibinfo{year}{1990}): \emph{\bibinfo{title}{Two classes of numbers
  appearing in the convolution of binomial-truncated Poisson and
  Poisson-truncated binomial variables}}.
\newblock {\slshape \bibinfo{journal}{Fibonacci Quarterly}}
  \bibinfo{volume}{28}(\bibinfo{number}{4}), pp. \bibinfo{pages}{321--333},
  \doi{10.1080/00150517.1990.12429470}.

\bibitemdeclare{article}{kyrvam}
\bibitem{kyrvam}
\bibinfo{author}{Andreas \surnamestart Kyriakoussis\surnameend} \&
  \bibinfo{author}{Malvina \surnamestart Vamvakari\surnameend}
  (\bibinfo{year}{2017}): \emph{\bibinfo{title}{Heine Process as a $q$-Analog
  of the Poisson Process: Waiting and Interarrival Times}}.
\newblock {\slshape \bibinfo{journal}{Communications in Statistics - Theory and
  Methods}} \bibinfo{volume}{46}, pp. \bibinfo{pages}{4088--4102},
  \doi{10.1080/03610926.2015.1078476}.

\bibitemdeclare{article}{malvina}
\bibitem{malvina}
\bibinfo{author}{Malvina \surnamestart Vamvakari\surnameend}
  (\bibinfo{year}{2020}): \emph{\bibinfo{title}{On Multivariate Discrete
  $q$-Distributions: A Multivariate $q$-Cauchy's Formula}}.
\newblock {\slshape \bibinfo{journal}{Communications in Statistics - Theory and
  Methods}} \bibinfo{volume}{49}, pp. \bibinfo{pages}{6080--6095},
  \doi{10.1080/03610926.2019.1626427}.

\end{thebibliography}

\end{document}